\documentclass[10pt,letterpaper,twocolumn]{article} %% two column, final layout

\usepackage{ol2}
\usepackage{amsmath}
\usepackage{amssymb}

\def\rtHz{\sqrt{\rm Hz}}

\begin{document}

\twocolumn[ 

\title{Optical Magnetometer Array for Fetal Magnetocardiography}

\author{Robert Wyllie,$^{1}$ Matthew Kauer,$^1$ Ronald T. Wakai,$^{2}$, and Thad G. Walker$^{1,*}$}

\address{
$^1$Department of Physics, University of Wisconsin-Madison, 1150 University Ave., Madison, WI 53706 USA\\
$^2$Department of Medical Physics, University of Wisconsin-Madison, 1111 Highland Ave., Madison, WI 53705 USA\\
$^*$Corresponding author: tgwalker@wisc.edu
}

\begin{abstract} We describe an array of spin-exchange relaxation free optical magnetometers designed for detection of fetal magnetocardiography (fMCG) signals.  The individual magnetometers are configured with a small volume with intense optical pumping, surrounded by a large pump-free region.  Spin-polarized atoms that diffuse out of the optical pumping region precess in the ambient magnetic field and are detected by a probe laser.  Four such magnetometers, at the corners of a 7 cm square, are configured for gradiometry by feeding back the output of one magnetometer to a  field coil to null uniform magnetic field noise at frequencies up to 200 Hz.  Using this array, we present the first measurements of fMCG signals using an atomic magnetometer.
\end{abstract}


 ] 

\noindent Spin-exchange relaxation free (SERF) magnetometers \cite{Kominis2003,Happer77} have comparable or better sensitivities as compared to SQUID devices, without the cryogenics.  These magnetometers have been configured as short baseline ($\sim 1$ cm) gradiometers and used for detection of  adult magnetoencephalography \cite{Xia2006meg,Johnson2010}.  A sub-cc SERF magnetometer was recently used for adult MCG and magnetorelaxometry \cite{knappe2010}.  We have recently demonstrated a high-sensitivity SERF magnetometer array for adult magnetocardiography (MCG) \cite{Wyllie11}.  A 25 detector array of conventional $M_x$ atomic magnetometers was recently demonstrated for adult MCG in a minimally shielded environment\cite{bison2009} but with insufficient sensitivity for fetal MCG (fMCG) detection.

For biomagnetic applications, cancellation techniques, such as gradiometry, are necessary to reduce environmental interference, even in  magnetically shielded rooms.  The optimum signal-to-noise ratio for gradiometry is attained when the gradiometer baseline is comparable to the source distance \cite{Pizzella2001}.  For fMCG, the fetal heart is typically located about 5-10 cm below the mother's skin, so a similar detector separation is desirable.  For such large detector separations, high quality SERF gradiometry with a single laser beam is challenging due to inevitable gradients in pumping rates that cause the frequency-dependence of the response to vary, giving imperfect cancellation of uniform magnetic fields at all frequencies.  A promising attack on this problem, demonstrated recently for a less sensitive $M_x$ magnetometer array, is to use active feedback from one or more channels to null the magnetic noise \cite{bison2009}.

In this Letter we demonstrate a 4-channel SERF magnetometer array with 7 cm channel spacing.  The individual channels are operated in a diffusive SERF regime, where the atomic precession is detected outside the spatially localized optical pumping regions.  Using this scheme, we demonstrate real-time  detection of  fMCG, a promising new method for diagnosing serious heart rhythm abnormalities in the fetus \cite{Hornberger2008}.  Signal processing from the 4 channels allows us to isolate the fMCG signal from the maternal MCG interference and to compare the waveform characteristics with those recorded on a commercial SQUID system.   Finally, in order to combat spatially uniform magnetic interference, we have operated the array  configured as a set of gradiometers using active cancellation of the magnetic field detected by one of the channels, and demonstrate interference rejection by 40dB.

\begin{figure}[b]
\centerline{\includegraphics[width=7.5cm]{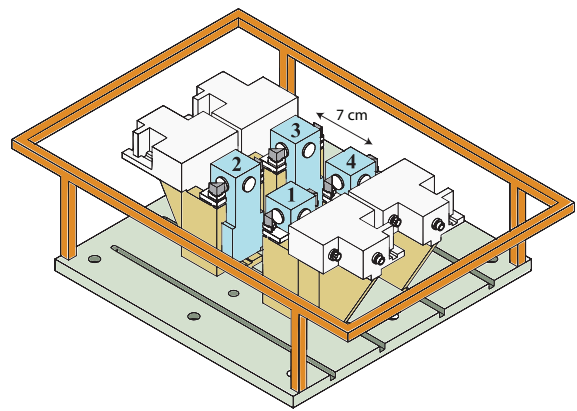}}
\caption{SERF biomagnetometer array.  Four magnetometers are symmetrically located in the plane of a single field coil.  The output $P_{1x}$ of one of the magnetometers is fed back to actively null $B_{1y}$.   Each channel consists of a heated glass magnetometer cell and a fiber-coupled laser/detection module.}\label{fig:array}
\end{figure}

The basic configuration of our SERF array is detailed in \cite{Wyllie11} and depicted in Fig.~\ref{fig:array}.
  Four magnetometers are located at the corners of a square, separated by  7 cm.  In each channel, a 4 mW circularly polarized 795 nm pump laser beam with a waist of 0.05 cm optically pumps the atoms in the $\hat{z}$-direction at the center of a 1 cm square glass cell.  Faraday rotation of a $\sim 780$ nm probe laser detects the spin-polarization $P_x$ along the $\hat{x}$-direction that is produced by magnetic fields along the $\hat{y}$ direction (perpendicular to the plane of the array).  When run as a hardware gradiometer, the signal $P_{1x}$ from channel 1 of the array is amplified by a gain stage and fed back to a current source driving a single rectangular magnetic field coil in order to keep $P_{1x}$=0.  The 4 channels are symmetrically located in the plane of the feedback field coil so that uniform magnetic fields are cancelled at each channel.  The signals in channels 2-4 are therefore gradiometric with respect to channel 1.

\begin{figure}[tb]
\centerline{\includegraphics[width=3.3 in]{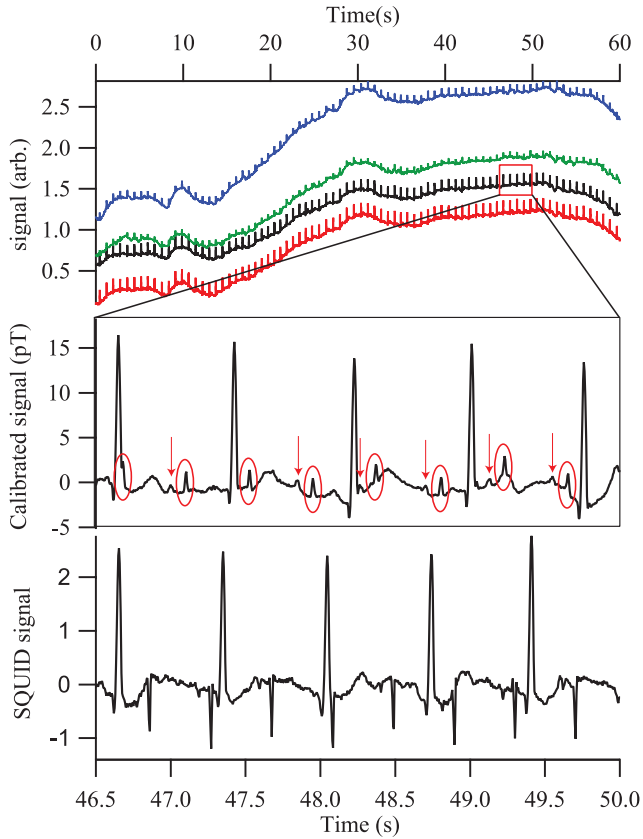}}
\caption{(top) Real-time magneto cardiogram taken from a pregnant mother and her 31 week fetus, showing all four magnetometer channels. (middle) Blown-up and calibrated portion of channel 2. The fetal heart QRS complexes are circled, and arrows identify the fetal P-wave components.  An 80 Hz low pass filter and a 60 Hz comb filter were applied.  (bottom) SQUID gradiometer signal, with the same filters applied.  The gradiometry suppresses the maternal MCG as compared to the fMCG. }\label{fig:realtime}
\end{figure}

Since the probe laser is much larger than the pump laser, and because the sensitivity is suppressed where the pump intensity is high, it primarily detects atoms that diffuse out of the pump beam region.  The cells contain 50 Torr of N$_2$ gas for excited-state quenching\cite{Lancor10c}, and roughly 20 Torr of He.  The diffusion coefficient is estimated to be $D=2.9$ cm$^2$/s.  At 150$^\circ$C, we estimate that the effective spin-relaxation rate of the atoms is about $\Gamma'=6.5$/s, including nuclear spin slowing-down effects\cite{Walker1997}.  Since the diffusion length $\Lambda=2\pi\sqrt{D/\Gamma'}=4.2$ cm is greater than the cell size, wall relaxation dominates the spin-relaxation in these cells.  The large probe beam and localized pump means we are primarily detecting atoms in regions with reduced AC Stark shifts and pump-induced relaxation.  Despite the relatively high wall relaxation rate in these cells, we thereby reach single channel noise levels of about 5 fT/$\rtHz$, only slightly above the noise levels of our shielded room.

We have used the SERF magnetometer array open loop to detect fetal magnetocardiography (fMCG) signals in real time.  The array is placed relatively centered over the fetus's position inside the mother's abdomen, with two of the channels relatively close to her heart, and the other two further away. 
Figure~\ref{fig:realtime} shows the raw signals observed from a fetus at 31 weeks gestation; the only filtering is a 80 Hz low-pass  filter and a 60 Hz comb filter.  The fetal QRS peaks are readily seen (circled in the blown-up portion of the signal) with sufficient signal-to-noise ratio to allow for the positioning of the detector to be adjusted to maximize the fetal signal in real time.  Note that the p-wave components, denoted by arrows, are also readily observed.  These are of particular importance for diagnosis of arrhythmias.

\begin{figure}[htb]
\centerline{\includegraphics[width=3.5 in]{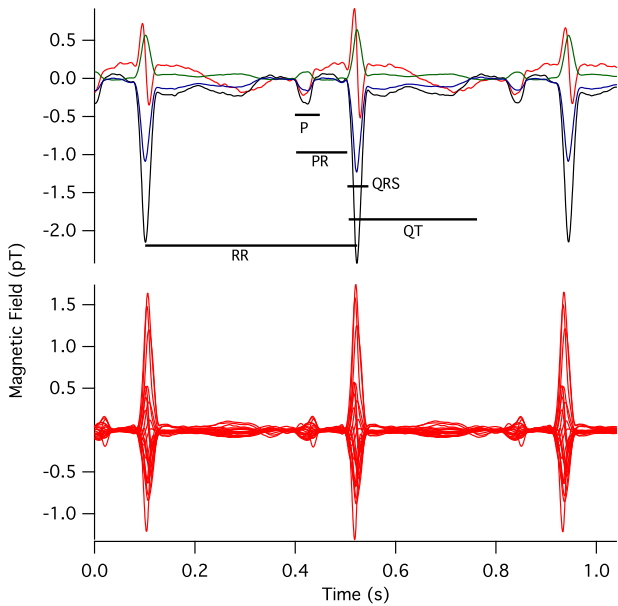}}
\caption{Comparison of the prototype optical magnetometer (top) and commercial SQUID (bottom)  signals, with timings between features corresponding to Table~\ref{timings}. }\label{fig:compare}
\end{figure}

The sensitivity of the raw fMCG tracings was similar for the SERF magnetometer and a 7-channel vector SQUID magnetometer (Tristan Vector Magnetometer, Tristan, Inc., San Diego) with 21 SQUID detectors.  The SQUID time series was acquired about 10 minutes after the SERF time series. A spatial filter \cite{Chen2001} was applied to isolate the fetal signal from the maternal interference, and averaged waveforms were computed using autocorrelation to time-align the fetal QRS complexes.  Fig. ~\ref{fig:compare} shows the averaged fMCG waveforms obtained using the two magnetometers and Table I shows a comparison of waveform interval measurements.  The intervals measured with the SERF and SQUID systems show excellent agreement.

\begin{table}[thb]
  \centering
  \caption{Intervals between various components of the fMCG waveforms, in msec. }
  \begin{tabular}{ccccccc} \\ \hline
    Method &RR &PR &P &QRS &QT& QTc\\ \hline
  SERF& 425 &95 &42 &47 &264 &405\\ \hline
 SQUID &415& 90& 41& 48& 241& 375 \\ \hline
 \end{tabular}\label{timings}
\end{table}

\begin{figure}[b]
\centerline{\includegraphics[width=3.5 in]{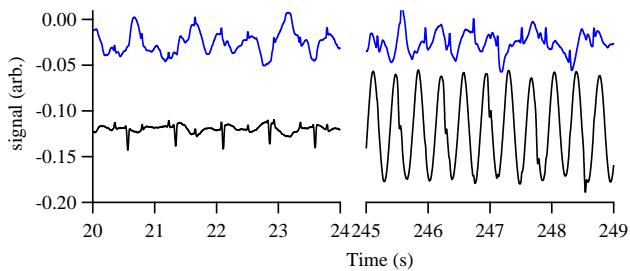}}
\caption{Output of two channels operating in the feedback mode.  The lower channel is the channel that is being fed back;  the signal displayed is the current supplied to the large magnetic field coil.  The upper trace shows the signal at another channel.  On the left is the data without a ventilation fan; the maternal signal happens to be nearly the same size at the two channels so the maternal signal is largely absent from the upper channel.  On the right, the ventilation fan is on.  It greatly affects the lower channel but not the upper one due to the feedback compensation.}\label{fig:feedback}
\end{figure}

In a hospital setting, large interfering background fields are often incompletely suppressed by the shielded room.  In our case, a nearby ventilation fan (turned off during the time series in Fig.~\ref{fig:realtime} ) is the largest interfering field.   By running the array with feedback from one of the channels, we have demonstrated real-time fMCG observation even in the presence of such interference.  Figure~\ref{fig:feedback}  shows the array run with feedback from channel 4.  The interfering field from the fan dominates even the maternal MCG field in channel 4.  Nevertheless, the feedback effectively cancels the interference in the other channels, and the fMCG signals are easily discerned.  The interference is suppressed in the other channels by 40 dB.

Although we have not yet done so,  it would be advantageous to run the other channels of the array with self-feedback in order to linearize the phase and amplitude response.  In this way, it should be possible to significantly improve the quality of the gradiometry.

Finally, we qualitatively discuss the diffusive SERF to illustrate its features.  At high alkali densities, SERF magnetometers are necessarily optically thick, producing substantial pumping rate and AC Stark gradients inside the cell.  Detecting atoms that have diffused out of a small pumping region allows reduced sensitivity to these effects.  Other technical advantages to this configuration include reduced optics size, ease of pumping several cells with the same beam, and less stringent demands on pump laser stability.

In conclusion, we have demonstrated that SERF magnetometers can be used for real-time fMCG detection and in an array can be used for spatial filtering of the maternal MCG signal for clinically interesting applications.  We expect that these results can be substantially improved upon by adding feedback to all channels, by increasing the channel count (either by using more magnetometers or by introducing detection of multiple field components in each magnetometer via a parametric modulation scheme \cite{Li2006parametric}), and by using vapor cells with better properties than those used here.  From our perspective, the pieces are in place to make high quality gradiometric SERF arrays with excellent sensitivity, at a cost per channel significantly lower than SQUIDs.

This work was supported by the NIH Eunice Kennedy
Shriver National Institute of Child Health \& Human Development, \#R01HD057965.  The authors are solely responsible for the content.  We appreciate help from Suhong Yu and Greg Smetana.

\bibliographystyle{ol}

\end{document}